# Prediction of Ideal Topological Semimetals with Triply Degenerate Points in NaCu$_3$Te$_2$ Family

Jianfeng Wang[1], Xuelei Sui[2,1], Wujun Shi[2], Jinbo Pan[3], Shengbai Zhang[4,1], Feng Liu*[5,6], Su-Huai Wei[1], Qimin Yan*[3] and Bing Huang*[1]

[1] *Beijing Computational Science Research Center, Beijing 100193, China*
[2] *Department of Physics and State Key Laboratory of Low-Dimensional Quantum Physics, Tsinghua University, Beijing 100084, China*
[3] *Department of Physics, Temple University, Philadelphia, PA 19122, USA*
[4] *Department of Physics, Applied Physics, and Astronomy, Rensselaer Polytechnic Institute, Troy, New York 12180, USA*
[5] *Department of Materials Science and Engineering, University of Utah, Salt Lake City, Utah 84112, USA*
[6] *Collaborative Innovation Center of Quantum Matter, Beijing 100084, China*

*Corresponding authors: Bing.Huang@csrc.ac.cn (B.H.); qiminyan@temple.edu (Q.Y.); fliu@eng.utah.edu (F.L.)

**Abstract**

**Triply degenerate points (TDPs) in band structure of a crystal can generate novel TDP fermions without high-energy counterparts. Although identifying ideal TDP semimetals, which host clean TDP fermions around the Fermi level ($E_F$) without coexisting of other quasiparticles, is critical to explore the intrinsic properties of this new fermion, it is still a big challenge and has not been achieved up to now. Here, we disclose an effective approach to search for ideal TDP semimetals via selective band crossing between antibonding *s* and bonding *p* orbitals along a line in the momentum space with $C_{3v}$ symmetry. Applying this approach, we have successfully identified the NaCu$_3$Te$_2$ family of compounds to be ideal TDP semimetals, where two and only two pairs of TDPs are located around the $E_F$. Moreover, we reveal an interesting mechanism to modulate energy splitting between a pair of TDPs, and illustrate the intrinsic features of TDP Fermi arcs in these ideal TDP semimetals.**

As topological phase extends from insulators [1,2] to semimetals [3-6], new quasiparticles analogous to elementary particles in high-energy physics emerge in these topological materials, such as Weyl (Dirac) fermions in Weyl (Dirac) semimetals [7-10]. Interestingly, the band theory has shown that the crystal symmetries in solids allow for the existence of other types of topological quasiparticle excitations even without high-energy counterparts [11], which can be hosted by three-, six-, or eight-fold degenerate points in the band structures [12]. Especially, the triply degenerate points (TDPs) [13-20], formed by the crossing of a double-degenerate band and a nondegenerate band, can be recognized as an intermediate phase between Weyl (double-degenerate) and Dirac (fourfold-degenerate) fermions. The TDP semimetals have been predicted to have some unique properties, e.g., Lifshitz transitions of Fermi surface [15,16], helical anomaly [16], large nonsaturating or negative magnetoresistance [21], and unconventional quantum Hall effects [22].

Generally speaking, the TDPs can appear along the high-symmetry lines with the $C_{3v}$ symmetry group in the Brillouin zone (BZ), because it allows for both one- (1D) and two-dimensional (2D) double-group representations. For example, the tensile-strained HgTe [13], CuPt-ordered InAs$_{0.5}$Sb$_{0.5}$ [14], WC-type or half-Heusler compounds [15-20] have been suggested as host candidates. Also, several experimental measurements have been carried out to reveal the electronic structures around TDPs in the MoP and WC compounds [23-25]. However, one of the key problems for exploring the intrinsic properties of TDP fermions is the lack of ideal TDP semimetals, in which the TDPs around the Fermi level ($E_F$) do not coexist with other quasiparticle bands. Therefore, it is of great importance to search for ideal host materials having only TDP fermions around $E_F$.

In this Letter, we disclose an effective approach to realize clean TDPs near the $E_F$ via selective band crossing between antibonding $s$ ($s^*$) and bonding $p$ orbitals along the line with $C_{3v}$ symmetry. Importantly, we have successfully identified that the NaCu$_3$Te$_2$ family of compounds are ideal TDP semimetals. Moreover, a simple mechanism has been revealed to control the energy splitting between the two adjacent TDP nodes. Finally, we illustrate the unique features of Fermi arc of TDP fermion, in comparison with 2- and 4-component fermions.

One of the most common characters found in previous TDP candidates is that the TDPs are mainly induced by different $d$ bands crossing near the Fermi level [15-19], e.g., crossing of $d_{x^2-y^2,xy}$-$d_{z^2}$ bands, as shown in Fig. 1(a). Because of the high degeneracy (5 orbitals) and localized nature (narrow band widths) of $d$ orbitals, these $d$ bands usually cross each other around $E_F$ multiple times in the entire BZ. As a result, besides the TDPs, other types of quasiparticle bands appear also around $E_F$ [15,17-19], which unfortunately overshadows the TDPs. Comparing with $d$ bands, $s$ and $p$ bands have low degeneracy (1 or 3 orbitals) and delocalized dispersion (wide band widths), which may play a useful role in creating clean TDPs. Here, we propose that the band inversion between antibonding $s^*$ orbital of cation at the conduction band minimum (CBM) and bonding $p$ orbitals of anion at the valence band maximum (VBM) in a compound may achieve clean $sp$-band TDPs, as illustrated in Fig. 1(a). When $s^*$ and $p$ bands cross each other along the $C_{3v}$ symmetric line in BZ, the $s^*$ bands will be double-degenerate ($J_z = \pm 1/2$) and $p_{x,y}$ bands will split into two nondegenerate bands ($J_z = \pm 3/2$) due to spin-orbit coupling (SOC) effect in a non-centrosymmetric structure. Belonging to different group representations, the hybridization between $s^*$ and $p_{x,y}$ orbitals is forbidden by symmetry, so that the $s^*$-$p$ band inversion will produce two pairs of desirable TDPs. Interestingly, in spite of the existence of $s^*$-$p$ band inversion in HgTe [13] and half-Heusler compounds [20], the higher symmetry of $p$ bands ($\Gamma_8$) at the $\Gamma$ point leads to multiple degenerated states coexisting with the TDPs that are resulted from the $p$-$p$ band crossing around $E_F$. Thus, the key to our approach is to find those compounds with the desired $s^*$-$p$ ($p_{x,y}$ here) band inversion in the whole BZ.

For a typical semiconducting compound with the $s^*$ orbital at the CBM and $p$ orbital at the VBM, their band energies can be determined from a two-level tight-binding model of the $s$-$s$ coupling and $p$-$p$ coupling between cation and anion, respectively [26]: $E_{\text{CBM}} = (\varepsilon_s^c + \varepsilon_s^a)/2 + \{[(\varepsilon_s^c - \varepsilon_s^a)/2]^2 + V_{ss}^2\}^{1/2}$, $E_{\text{VBM}} = (\varepsilon_p^c + \varepsilon_p^a)/2 - \{[(\varepsilon_p^c - \varepsilon_p^a)/2]^2 + V_{pp}^2\}^{1/2}$. The $\varepsilon_s^c$ and $\varepsilon_s^a$, $\varepsilon_p^c$ and $\varepsilon_p^a$ are the cation and anion $s$ and $p$ atomic orbital energies, respectively, and $V_{ss}$ and $V_{pp}$ are the coupling potentials for $s$ and $p$ states, respectively. First, to get band inversion between the $s^*$- and $p$-orbital bands, a simple way is to find a material with close energies between $E_{CBM}$ and $E_{VBM}$. Considering the suitable values of $\varepsilon_s^c$ and $\varepsilon_p^a$, together with the typical strengths of $V_{ss}$ and $V_{pp}$ [26], we can confine our search in compounds with cation and anion candidates listed in Fig. 1(b). Besides the $s$-$s$ and $p$-$p$ couplings, other orbital hybridizations may also affect the energies of the $s$ and $p$ bands, which are compound dependent [27]. Second, to acquire both 1D and 2D double group representations for the TDPs, our search is further confined into those compounds with the $C_{3v}$ subgroup. Following these rules, we have successfully identified that the NaCu$_3$Te$_2$ is the targeted TDP semimetal [28].

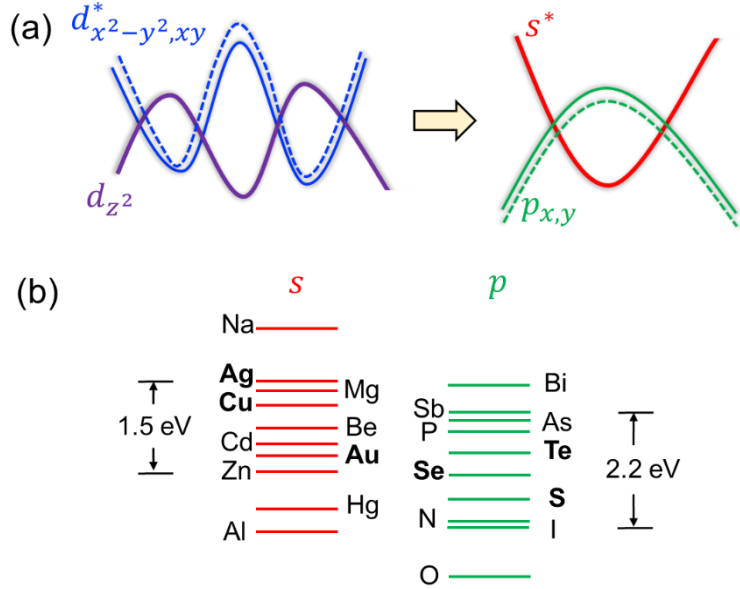

Figure 1. (a) Schematic plot of $d^*$-$d$ TDP semimetals and our strategy for $s^*$- $p$ TDP semimetals. Double-degenerate bands are drawn as thick solid lines, whereas nondegenerate bands are drawn as thin solid or dashed lines. (b) Atomic $s$ and $p$ orbital energy levels of targeted cations and anions considered in our study.

As shown in Fig. 2(a), NaCu$_3$Te$_2$ (ICSD No.: 60860) has a non-centrosymmetric rhombohedral structure with space group $R3m$ (No. 160). The fully relaxed lattice constant for its 18-atom conventional cell is $a = b = 4.25$ Å, $c = 23.11$ Å, consistent with the experimental values ($a = b = 4.276$ Å, $c = 23.78$ Å) [29]. Its structure can be visualized in terms of a cubic-close-packed array of Te atoms, with Na and Cu occupying alternatingly the interstitial layers. Na is in an octahedral coordination with an average Na-Te bond length of 3.11 Å, and a small shift occurs for Na from the center of octahedral site towards the Te$_2$ atom. Cu atoms occupy the tetrahedral and octahedral voids with a small displacement from the centers of these sites. Cu$_1$ and Cu$_2$ are in tetrahedral coordination with an average Cu-Te bond length of 2.717 Å and 2.736 Å respectively. Cu$_3$ is in octahedral coordination with a large shift towards Te$_1$ [28].

By calculating band structure in the entire BZ, as shown in Fig. 2(b), we can find that NaCu$_3$Te$_2$

is an ideal TDP semimetal with the desired $s^*$-$p$ band inversion solely along the $\varGamma Z$ line that has the $C_{3v}$ symmetry, as shown in Fig. 2(c). Along the $\varGamma Z$ line, the $C_{3v}$ symmetry group has one 2D ($\Lambda_6$) and two 1D ($\Lambda_4$, $\Lambda_5$) double-group representations. Including SOC effect, the $s^*$ band ($\Lambda_6$ representation) belongs to double-degenerate $J_z = \pm 1/2$ states, whereas the $p$ band ($p_{x,y}$ here, $\Lambda_4$ and $\Lambda_5$ representations) splits into two nondegenerate $J_z = \pm 3/2$ states, as demonstrated in left panel of Fig. 2(d). Consequently, the band crossing near $E_F$ generates a pair of TDPs along $Z\varGamma$, as shown in the middle panel of Fig. 2(d). Along the direction perpendicular to $Z\varGamma$ line, each TDP will split into three nondegenerate bands, as revealed in the right panel of Fig. 2(d). Thus, the TDPs are strictly protected by the $C_{3v}$ symmetry. As expected from Fig. 1(b), along the other $\varGamma Z$ line there is another pair of identical TDPs. The position of these two pairs of TDP in the momentum space are (0, 0, $\pm 0.0943$ Å$^{-1}$) and (0, 0, $\pm 0.0924$ Å$^{-1}$), respectively. The topological nature of TDPs in NaCu$_3$Te$_2$ is further confirmed by calculating the $Z_2$ topological invariants, which are well-defined in both the $k_z = 0$ ($Z_2 = 1$) and $k_z = \pi$ planes ($Z_2 = 0$) [28].

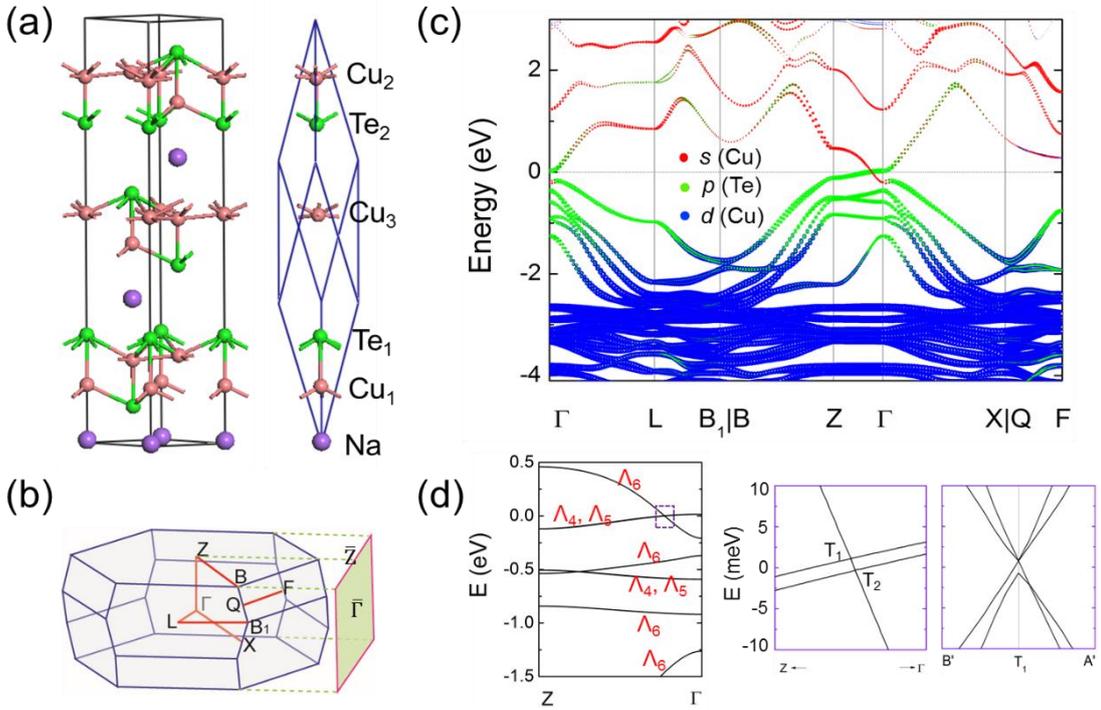

Figure 2. (a) Conventional unit cell (black line) and primitive cell (blue line) of NaCu$_3$Te$_2$. In the primitive cell, two nonequivalent Te atoms are labeled as Te$_1$ and Te$_2$ while three nonequivalent Cu atoms are labeled as Cu$_1$, Cu$_2$ and Cu$_3$, respectively. (b) BZ of primitive cell of NaCu$_3$Te$_2$ and its projection towards (010) surface. The red lines in BZ depict the high-symmetry lines. (c) Band structure (with SOC effect) with different atomic orbitals projections. (d) Left panel: band structure along $Z\varGamma$ with the labels of three double-group representations. Middle and right panels: band structures around $E_F$ along and perpendicular to the $Z\varGamma$ direction, respectively. Two TDPs are labeled by T$_1$ and T$_2$ points.

To understand the origin of TDPs, we start from the atomic energy levels and consider the effects of orbital hybridization, crystal-field splitting, and SOC on the band evolution in the vicinity of $\varGamma$ point, as shown in Fig. 3(a). Without hybridizations, the Cu $s$, Te $p$ and Cu $d$ orbitals are close in energy. With hybridizations and crystal field splitting (stage I), the antibonding $s^*$ and bonding $p$ bands are formed, and the crystal-field effect makes the $p_z$ orbital split from the double-degenerated $p_{x,y}$ orbitals. The additional strong $p$-$d$ hybridizations upshift (downshift) the $p$ ($d$) orbitals to higher (lower) energy positions. Among Te$_1$ and Te$_2$ atoms, the Te$_2$ atom has a stronger $p$-$d$ hybridization effect due to the shorter Te$_2$-Cu bond lengths. Consequently, the Te$_2$ $p$ orbitals will be pushed to higher energy levels [green solid lines in Fig. 3(a)] than that of Te$_1$ $p$ orbitals [green dashed lines in Fig. 3(a)]. In the vicinity of $\Gamma$ point, the Te$_2$ $p_{x,y}$ orbitals are pushed up to an even higher energy position than that of Cu $s^*$ orbital, while at all other high-symmetry points all the Te $p$ orbitals still have lower energies than that of Cu $s^*$ orbital, which give rise to a band inversion solely around the $\Gamma$ point in the entire BZ. In stage II, the SOC effect mixes spin and orbital angular momenta while preserving the total angular momentum. The $p_{x,y}$ orbitals further split into $J_z = \pm 3/2$ and $J_z = \pm 1/2$ states, meanwhile both $s$ and $p_z$ orbitals evolve into $J_z = \pm 1/2$ states. Around the $\Gamma$ point, the (Te$_2$) $J_z = \pm 3/2$ states are still located above the $s^*$-type $J_z = \pm 1/2$ states. Along $\Gamma Z$, all $J_z = \pm 1/2$ states belong to the $\Lambda_6$ representation and $J_z = \pm 3/2$ states belong to the $\Lambda_4$ and $\Lambda_5$ representations. For the non-centrosymmetric system, the SOC effect can further lift the degeneracy of $J_z = \pm 3/2$ states, and as a result the band crossing of $J_z = \pm 3/2$ and $J_z = \pm 1/2$ states along $\Gamma Z$ can generate two ideal TDPs in the whole BZ.

Since the energy splitting between the two TDPs along $\Gamma Z$ in NaCu$_3$Te$_2$ is mostly contributed by the energy splitting of $J_z = \pm 3/2$ states, it is therefore crucial to develop understanding on what modulate the size of energy splitting between the $J_z = \pm 3/2$ ($\Lambda_4$ and $\Lambda_5$) states, which may be important for future applications. We discover that the splitting of $\Lambda_4$ and $\Lambda_5$ states is mostly contributed by the Dresselhaus SOC effect [30], which is proportional to the momentum, $\Delta E = C_k k$, where $k$ is along the direction with $C_{3v}$ symmetry and $C_k$ determines the size of splitting between the $\Lambda_4$ and $\Lambda_5$ states. The $C_k$ originates from the second-order interaction between the $J = 3/2$ states and the uppermost cation $d$ core levels in the spin-orbit operator $H_{so}$ [31]. It can be deduced $C_k = \alpha \Delta_d S \beta / [E(3/2) - E_d]$, where $\alpha$ is a constant, $\Delta_d$ is the spin-orbit splitting of the $d$ orbitals of the cation with $E_d$ of its energy, $E(3/2)$ is the valence band energy with $J = 3/2$ states, $\beta$ is the admixture coefficient of $d$ orbitals in the valence band, and $S$ is the corresponding matrix element of momentum $p$ [$S = i \langle 3/2 | p_x | d_{x^2 - y^2} \rangle$]. Therefore, the $p$-$d$ hybridization, allowed only if the inversion symmetry is broken, together with spin-orbit splitting of $d$ orbital of Cu, determines the splitting magnitude between $\Lambda_4$ and $\Lambda_5$ states in NaCu$_3$Te$_2$.

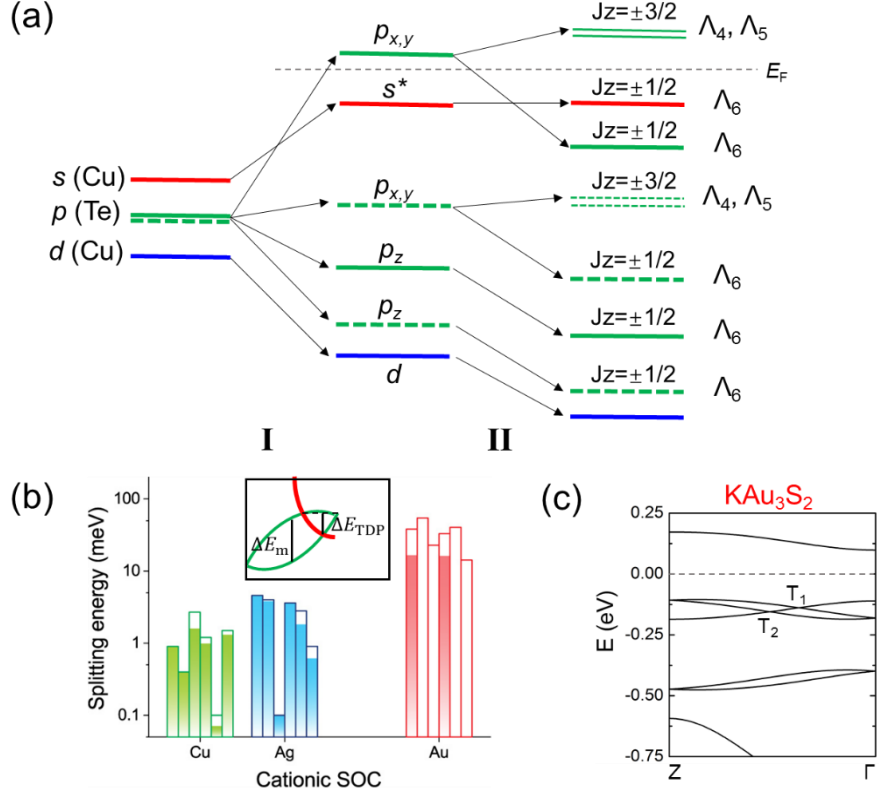

Figure 3. (a) Schematic diagram of band evolution in the vicinity of $\Gamma$ point: stage I represents the orbital hybridization and crystal-field splitting effect, and stage II represents SOC effect (see text). (b) The maximum splitting energy ($\Delta E_\mathrm{m}$) between two $J_z = \pm 3/2$ states (total column height) and the realistic splitting energy ($\Delta E_\mathrm{TDP}$) between two TDPs (solid column height) for $AB_3X_2$. $\Delta E_\mathrm{m}$ and $\Delta E_\mathrm{TDP}$ are illustrated in the inset, where band crossing occurs between $J_z = \pm 1/2$ (red) and $J_z = \pm 3/2$ (green) states. The NaCu$_3$Te$_2$ family are divided into three groups as $B$ = Cu, Ag, Au. For each group, six columns represent Na$B_3X_2$ ($X$ = S, Se, Te) and K$B_3X_2$ ($X$ = S, Se, Te) from left to right in sequence. (c) Band structure of KAu$_3$S$_2$ along $Z\Gamma$. The $E_\mathrm{F}$ is set to zero.

To confirm this, we have extended NaCu$_3$Te$_2$ to its family compounds $AB_3X_2$ ($A$ = Na, K; $B$ = Cu, Ag, Au; $X$ = S, Se, Te) by isovalent cation/anion replacements. These $AB_3X_2$ compounds could have similar stable structure to that of NaCu$_3$Te$_2$ according to our formation energy calculations [28]. Interestingly, some materials are intrinsic TDP semimetals while the others are semiconductors that need additional charge doping to shift the $E_\mathrm{F}$ to TDPs [28]. In all the $AB_3X_2$ compounds, the bands along $\Gamma Z$ have similar characteristics as that of NaCu$_3$Te$_2$, i.e., the $p_{x,y}$ orbitals split into double-degenerate $J_z = \pm 1/2$ states and two nondegenerate $J_z = \pm 3/2$ states with SOC effect, and the splitting magnitude between these two $J_z = \pm 3/2$ states is determined by $C_k$. In Fig. 3(b), we have classified the maximum splitting energy ($\Delta E_\mathrm{m}$) between $J_z = \pm 3/2$ states in $AB_3X_2$ materials into three sub-groups in terms of the SOC strength of cation $B$ (Cu, Ag, Au). Each sub-group has six materials, i.e., Na$B_3X_2$ ($X$ = S, Se, Te) and K$B_3X_2$ ($X$ = S, Se, Te). Generally, our calculations confirm that the SOC strength of $B$ element will overall determine the maximal splitting of $J_z = \pm 3/2$, while the zero-order SOC of $X$ anion has almost no impact on the energy splitting. Interestingly, in each sub-group the diversity of $\Delta E_\mathrm{m}$ of $AB_3X_2$ ($A$ = Na, K; $X$ = S, Se, Te) as a function of $A$ and $X$ is contributed by the distinct $p$-$d$ hybridization strength, i.e. the stronger the $p$-$d$ hybridization, the larger the splitting between $J_z = \pm 3/2$ [28]. Another way to tune the *p-d* hybridization strength is strain (pressure) engineering. For example, it is found that compressive (tensile) strain can increase (decrease) $\Delta E_\mathrm{m}$ by increase (decrease) *p-d* hybridization [28]. Thus, our calculations not only verify our model, but also provide an effective way to modulate the energy splitting of $J_z = \pm 3/2$ states. It is also note that the size of $\Delta E_\mathrm{m}$ is not guaranteed to be the realistic energy splitting of two TDPs ($\Delta E_\mathrm{TDP}$), as the latter also depends on the position of band crossing between the $J_z = \pm 3/2$ and $J_z = \pm 1/2$ states, i.e., $\Delta E_\mathrm{TDP}$ can be smaller than $\Delta E_\mathrm{m}$, as shown in Fig. 3(b). Generally, by the choices of specific $AB_3X_2$, $\Delta E_\mathrm{TDP}$ can be dramatically tuned from several meV (such as NaCu$_3$Te$_2$) to dozens of meV [such as KAu$_3$S$_2$ in Fig. 3(c)].

NaCu$_3$Te$_2$ family can serve as ideal platforms to study the unique surface states and Fermi arcs of TDP semimetals. Fig. 4(a) shows the surface projected band for the (010) surface of a semi-infinite NaCu$_3$Te$_2$ system. There is a clear Dirac cone-like surface state centering at $\bar{\Gamma}$. The upper branch and lower branch connect to the conduction and valence bands, respectively. Along $\bar{\Gamma}\bar{X}$, the number of crossings between surface states and any in-gap energy level is odd, confirming the nontrivial $Z_2$ in the $k_z = 0$ plane. Along $\bar{\Gamma}\bar{Z}$, the two branches of topological surface states terminate at two bulk TDPs respectively. The Fermi surface with $E_F$ shifting to -10 meV is shown in Fig. 4(b). Around $\bar{\Gamma}$, two pieces of half-circle Fermi arcs touch with each other near the surface projection of one TDP (T$_1$). Outside these two connected Fermi arcs, another branch of Fermi arc begins or ends at an adjacent point where the other TDP (T$_2$) projection is located, and it merges into the valence band states. Notably, we find that the separation of these two singularity points emitting different branches of Fermi arcs is related to the splitting of two adjacent TDPs.

From the ideal Fermi arc discovered in NaCu$_3$Te$_2$, the distinct features of Fermi arc of TDPs with Dirac and Weyl points can be understood in Fig. 4(c). For Dirac points that are resulted from crossing of two double-degenerate bands, the two Fermi arcs touch at the projections of Dirac points, and they form a closed circle with a discontinuous Fermi velocity (upper panel). The two Weyl points with opposite chirality can be viewed as the split of one Dirac point, and the split is moved away from the high-symmetry line, so that the Fermi arcs of Weyl points are disconnected (middle panel). In contrast, the two adjacent TDPs are split along the high-symmetry line, hence their Fermi arcs are separated along this high-symmetry line (bottom panel). When the $E_F$ crosses one TDP with higher energy (T$_1$), two pieces of Fermi arcs touch at the projection of T$_1$; when the Fermi level crosses the other TDP with lower energy (T$_2$), another branch of Fermi arc appears and merges into the valence band [28]. The split of two TDPs ($\Delta k_\mathrm{TDP}$ or $\Delta E_\mathrm{TDP}$) will influence the separation of these two branches of Fermi arcs. It is expected that the intrinsic characteristics of TDP fermions we discovered here can be experimentally confirmed in the future.

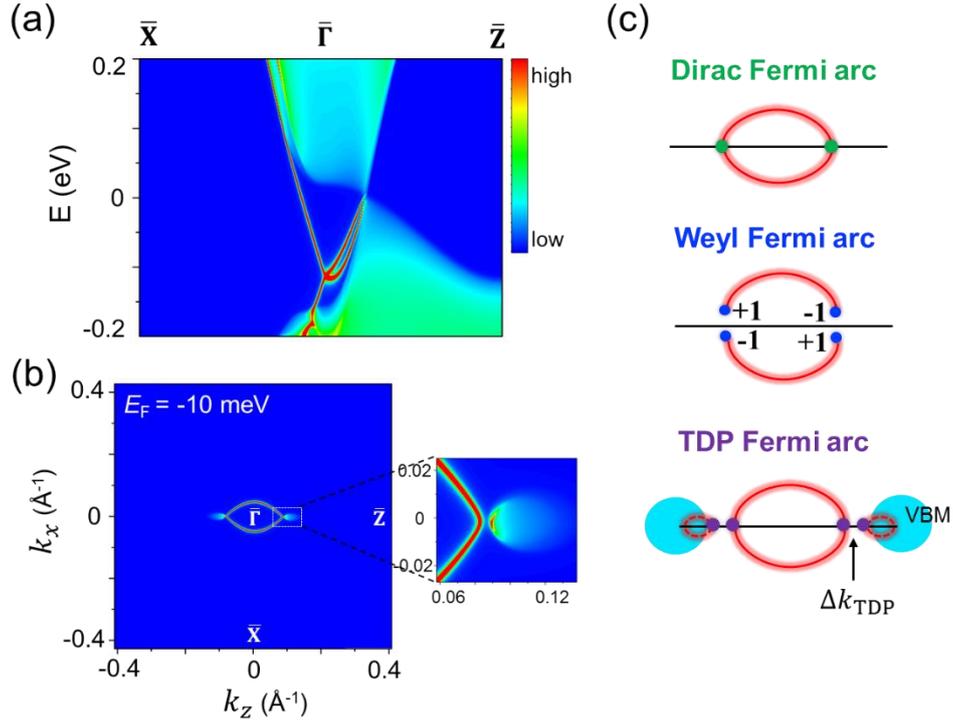

Figure 4. (a) and (b) Surface projected bands and Fermi surface for the (010) surface of a semi-infinite NaCu$_3$Te$_2$ system. (c) Comparison of Fermi arcs between Dirac, Weyl and TDP fermions.

In conclusion, we disclose an effective approach to search for ideal TDP semimetals. We further discovered that the NaCu$_3$Te$_2$ family of compounds are ideal TDP semimetals with unique Fermi surface states and Fermi arcs. We also revealed an effective mechanism to modulate the energy splitting between the two adjacent TDP nodes in these materials.

**Acknowledgements:** J. W., S.-H. Wei, and B. H. acknowledge the support from NSFC (Grant No. 11574024) and NSAF U1530401. F. L. acknowledges the support from US-DOE (Grant No. DE-FG02-04ER46148). S. B. Z. acknowledges the support from US-DOE (Grant No. DE-SC0002623). Part of the calculations were performed at Tianhe2-JK at CSRC.